\documentclass[useAMS]{mn2e}
\usepackage{epsfig}
\usepackage{color}
\newcommand{\alfosc}{{\sc alfosc}}

\newcommand{\kms}{km\,s$^{-1}$}

\newcommand{\CIII} {C\,{\sc iii}}

\newcommand{\FeII} {Fe\,{\sc ii}}
\newcommand{\HeI}  {He\,{\sc i}}
\newcommand{\HeII} {He\,{\sc ii}}

\newcommand{\NaI}  {Na\,{\sc i}} 
\newcommand{\NIII} {N\,{\sc iii}}

\newcommand{\OIII} {O\,{\sc iii}}

\newcommand{\Lyalpha}{Ly$\alpha$}
\newcommand{\Halpha} {H$\alpha$}
\newcommand{\Hbeta}  {H$\beta$}
\newcommand{\Hgamma} {H$\gamma$}

\begin{document}

\title[Multi-epoch Spectroscopy of IY\,UMa]
{Multi-epoch Spectroscopy of IY\,UMa: Quiescence, Rise, Normal Outburst \& Superoutburst}
\author[D. J. Rolfe et al.]
  {Daniel J. Rolfe$^{1,2}$, Carole A. Haswell$^1$, Timothy M. C. Abbott$^{3,4}$
\newauthor Luisa Morales-Rueda$^{5,6}$, T. R. Marsh$^{5,7}$ and G. Holdaway$^{5,8}$\\
$^1$Department of Physics \& Astronomy, The Open University, Walton Hall, Milton Keynes, MK7 6AA, UK.\\
$^2$Department of Physics \& Astronomy, University of Leicester, University Road, Leicester, LE1 7RH, UK.\\
$^3$Cerro Tololo Inter-American Observatory, Casilla 603, La Serena, Chile\\
$^4$Nordic Optical Telescope, Roque del Los Muchachos \& Santa Cruz de La Palma, Canary Islands, Spain\\
$^5$Department of Physics \& Astronomy, Southampton University, Southampton, SO17 1BJ, UK.\\
$^6$Department of Astrophysics, University of Nijmegen,
P.O. Box 9010, 6500 GL Nijmegen, The Netherlands.\\
$^7$
Department of Physics, University of Warwick, Coventry, CV4 7AL, UK.\\
$^8$
Fault Studies \& Fuel Branch, British Energy Generation Ltd, Gloucester, GL4 3RS\\
}
\date{Accepted. Received}
\pagerange{\pageref{firstpage}--\pageref{lastpage}}
\pubyear{2002}
\maketitle \label{firstpage}
\begin{abstract}
We exploit rare observations covering the time before
and during a normal outburst  
in the deeply-eclipsing SU\,UMa
  system IY\,UMa to study the dramatic changes in the accretion flow
  and emission at the onset of outburst.  Through Doppler tomography
  we study the emission distribution, revealing classic accretion flow
  behaviour in quiescence, with the stream-disc impact ionizing the
  nearby accretion disc. We observe a delay of hours to a couple of
  days between the rise in continuum and the rise in the emission
  lines at the onset of the outburst. From line profiles and Doppler
  maps during normal and superoutburst we conclude that reprocessing
  of boundary layer radiation is the dominant emission line mechanism
  in outburst, and that the normal outburst began in the outer disc.
  The stream-disc impact feature (the `orbital hump') 
  in the \Halpha\ line flux light curve disappears before the 
  onset of the
  normal outburst, and may be an observable signal heralding an 
  impending outburst. 
\end{abstract}


\section{Introduction}

IY\,UMa is an SU\,UMa type dwarf nova cataclysmic variable (CV). These
are systems in which a white dwarf accretes via Roche lobe overflow
from the donor, forming an accretion disc around the white dwarf which
undergoes a series of short normal outbursts ($\sim 5$ days for
IY\,UMa), and longer superoutbursts ($\sim 3$ weeks). The
  outbursts are thought to result from the switching of the disc
  between a cool, neutral, low-viscosity state 
%where the hydrogen is neutral
  and a hot, ionised viscous state. In the cool state very little mass
  is transferred inwards, while in the hot state mass is rapidly
  accreted onto the white dwarf. Superoutbursts are thought
to differ from normal outbursts due to the onset of a tidal
instability in the disc, in which a 3:1 resonance between tidal forces
and orbits of particles in the disc leads to a distorted, precessing
disc.  Periodic luminosity modulations, called superhumps, with
periods a few per cent longer than the orbital period are the defining
characteristic of superoutbursts. Superhumps result from
  the interaction of the donor orbit with the eccentric disc. For a
review see \nocite{Osaki:1996}Osaki (1996) and for a recent discussion
\nocite{OsakiMeyer:2003}Osaki \& Meyer (2003).

Patterson et al.  (2000, hereafter P2000) and \nocite{RolfeEt:2001a}Rolfe, Haswell \&  Patterson (2001b)
present detailed photometric studies of IY\,UMa's January 2000
superoutburst and superhumps.  Spectroscopic observations reveal a
bright hotspot and deep white dwarf absorption lines during quiescence
\nocite{RolfeEt:2001b,PattersonEt:2000}(Rolfe, Abbott \&  Haswell 2001a; Patterson {et~al.} 2000); an eccentric disc during
superoutburst \nocite{WuEt:2001}(Wu {et~al.} 2001); lines powered by reprocessed boundary
layer emission during outburst \nocite{RolfeEt:2002b}(Rolfe, Abbott \&  Haswell 2002a) and an M dwarf
donor star \nocite{RolfeEt:2002a}(Rolfe, Abbott \&  Haswell 2002b).

We present spectroscopic observations of IY\,UMa in
quiescence, rise, in outburst and in superoutburst. 
%{\sout{We discuss the interesting features
%of the detailed quiescent observations, in particular the spectrum of
%the stream-disc impact region. We use spectrophotometry of
%  the period immediately preceding and during a normal outburst to
%  understand the mechanism powering emission.} This is further backed
%up by our spectroscopy of a superoutburst.}} 
Section 2 presents the observations.
In Section \ref{AvSpec} we
discuss the average spectra epoch by epoch. We cover the lightcurves
in Section \ref{Lightcurves} and the line profile variations and
detailed accretion flow in Section \ref{Tomography}.
In Sections 6 and 7 we discuss and summarise our findings.

\section{Observations and data reduction}

\begin{figure}
\centerline{\epsfig{file=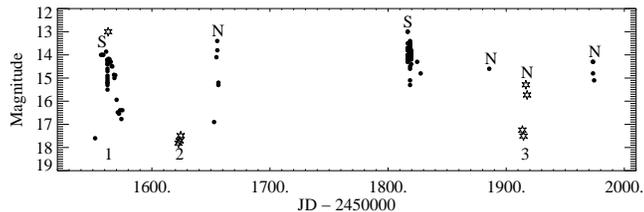,width=\columnwidth}}
\caption{  
  VSNET observations of IY\,UMa outbursts (circles) and observing runs
  presented in this work (stars). 'S' denotes a superoutburst, 'N' a
  normal outburst. The observations are marked: 1 - WHT, January 2000,
  2 - NOT, March 2000 and 3 - NOT, January 2001.}
\label{ObsFig}
\end{figure}

Fig. \ref{ObsFig} marks our observations in relation to the longterm
lightcurve from the VSNET network.

Orbital phases for all observations in this work were calculated
using the white dwarf ephemeris in P2000.

\subsection{WHT}

On 2000 January 19 we used
%We obtained spectroscopy of IY\,UMa on the 19th of January 2000 using
the ISIS double beam spectrograph 
%{\sout{mounted}} 
on the 4.2\,m William
Herschel Telescope (WHT) on La Palma
%Details of the instruments and
%observations in each arm are given in 
(see Table \ref{ObsTable} for details) 
%The setup
%allowed high-resolution 
to monitor 
%of 
\Halpha\ and \HeI\ 
6678\AA\ in the red arm and \Hgamma, \Hbeta, \HeII\ and several
He\,{\sc i} lines in the blue arm.

Standard CCD processing, optimal extraction of spectra
%After debiasing and flat-fielding the frames by bias and tungsten lamp
%exposures, spectral extraction proceeded according to the optimal
%algorithm of 
\nocite{Marsh:1989}(Marsh 1989),  
%The data were 
wavelength calibration and
%using CuAr and CuNe arc lamps and corrected for
instrumental response and extinction 
corrections were performed.
%using the flux standard HD19445
%\nocite{OkeGunn:1983}( ). The arcs were extracted using the profile
%determined to extract their associated image to avoid possible
%systematic errors caused by tilted spectra. 
Uncertainties on every
point were propagated through every stage of the data reduction.

As we did not put a comparison star in the slit with IY\,UMa the spectra
have not been corrected for slit losses. This means that the
variability seen from one spectrum to the next could be due to passing
clouds and it is most probably not due to the intrinsic variability of
the system.

%In this paper we will use the terms red and blue WHT spectra referring to
%the spectra taken with the red and blue arms of ISIS repectively.

\subsection{NOT}

The Nordic Optical Telescope (NOT) observations comprise 
%consist of 
about 14 binary orbits of spectra from 2000 March and 2001 January 
taken using
the \alfosc\ spectrograph 
%using grisms 6, 7 and 8, as detailed in
(see Table \ref{ObsTable}).

\begin{table}
  \caption{The observations. $\Delta\lambda$ is the FWHM spectral 
resolution 
%of the spectra measured from arc lines. 
$N$ is the number of spectra. The instrument
column lists the grating, CCD and readout mode for the red and blue
ISIS arms in WHT observations and the grism number for the NOT
observations.}

\label{ObsTable}
\begin{center}
\begin{tabular}{|c|c|c|c|c|c|c|}
\hline

Tel.                 & Instr. & Orbits & N   & Exp. & $\lambda$~range & $\Delta\lambda$ \\
Date                 &       &        &     &    (s)    & (\AA)           & (\AA)           \\
\hline
WHT        & ISIS   &        &     &           &                 &                 \\
19/1/00    & R1200R    &  1.4   & 64  &       80  & 6350--6750      &     0.8         \\
           & Tek/Quick    &        &     &           &                 &                 \\
           &  R1200B   & 1.3   & 53  &       80  & 4270--5070      &     0.9         \\
           & EEV/Std.    &        &     &           &                 &                 \\
\hline
NOT        & ALFOSC &        &     &           &                 &                 \\
18/3/00    & 6&  4.1   & 59  &      300  & 3180--5550      &     6           \\
19/3/00    & 7&  2.9   & 80  &      180  & 3820--6840      &     6           \\
20/3/00    & 7&  1.6   & 64  &      120  & 3820--6840      &     5           \\
3/1/01     & 8&  1.6   & 21  &      360  & 5810--8350      &     5           \\
4/1/01     & 8&  1.2   & 18  &      360  & 5810--8350      &     5           \\
6/1/01     & 7&  1.1   & 58  &       60  & 3820--6840      &     5           \\
7/1/01     & 7&  2.2   & 131 &       60  & 3820--6840      &     5           \\
\hline
\end{tabular}
\end{center}
\end{table}

The 2000 March observations were reduced and flux calibrated
following standard procedures
%using IRAF 
without corrections for slit losses. 
%Flux standard
%observations were used to correct the spectra for atmospheric
%extinction and the instrumental response. 
The March 18th spectra are
reliable only between about 4000\AA\ and 5200\AA\ due to wavelength
calibration problems.

The 2001 January exposures of IY\,UMa included a nearby comparison
star for correction of slit losses. Wide slit exposures of IY\,UMa and
the comparison, and of a flux standard, were taken and used
for flux and slit
loss calibration. 
%Arc lamp exposures were taken at each telescope
%position. Various halogen lamp and sky flats were also taken.

The 2001 Jan observations were fully flux calibrated and slit
loss corrected above $\sim$4800\AA\
%footnote{\color{red}The comparison star was
%  too faint for reliable calibration below this wavelength in many
%  exposures.}, 
and degraded only by the wavelength dependent slit
losses at shorter wavelengths. 
%The blue ends of the grism 8
%observations discussed in this paper were also flux calibrated and
%slit loss corrected.

\section{Average spectra}
\label{AvSpec}

\subsection{2000 March: IY\,UMa in quiescence}
\label{AvSpec2000}

\begin{figure}
\centerline{\epsfig{file=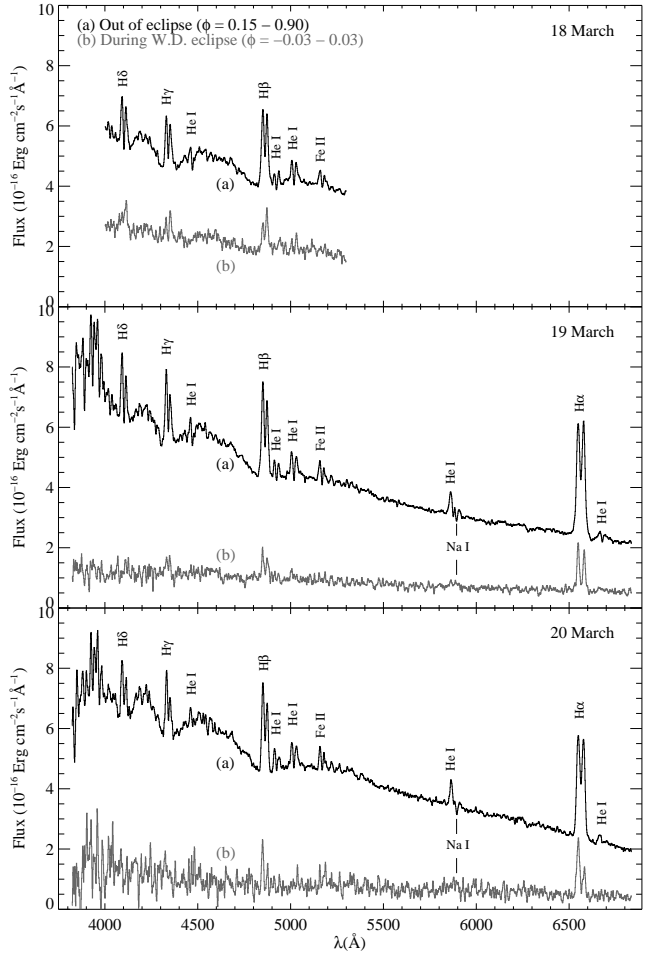,width=\columnwidth}}
\caption{
  Average spectra of IY\,UMa during quiescence in March 2000.}
\label{AvSpectraFig2000}
\end{figure}

Fig. \ref{AvSpectraFig2000} shows 2000 March average nightly spectra 
during and outside eclipses
%when neither 
%hotspot
%and white dwarf eclipse
%is eclipsed 
(phase ranges given in Fig. \ref{AvSpectraFig2000}). 
The spectra have been smoothed by slightly less than the
instrumental resolution.
Strong, double-peaked Balmer, \HeI\ and \FeII\ emission lines are
superimposed on a blue continuum. Broad absorption wings are seen
around the Balmer emission lines except H$\alpha$ 
%{\color{red}
and
there are deep cores between the double peaks of the lines, reaching
below the continuum in some cases. 
%\sout{The depth of the core between
% the Balmer double peaks is greatest for the higher order lines,
%  being below the continuum for H$\gamma$ and H$\delta$. The \HeI\ and
%  \FeII\ lines also show deep cores between their double peaks,
%  reaching below the continuum.}} 
Such features are seen in two other
high inclination dwarf novae in quiescence - Z\,Cha
\nocite{MarshEt:1987}(Marsh, Horne \& Shipman 1987) and OY\,Car \nocite{HessmanEt:1989}(Hessman {et~al.} 1989). Weak \NaI\,
doublet absorption at 5890--5896\AA\ is superimposed on \HeI\ 5876\AA.

The full width of the Balmer absorption wings is about 20,000\,\kms,
far too large to be Doppler-shifted disc material.
%\nocite{MarshEt:1987}Marsh {et~al.} (1987) concluded that the corresponding features in
%Z\,Cha are Stark broadened in the dense region close to the white
%dwarf. 
%Fig. \ref{AvSpectraFig2000} shows that 
The broad absorption
wings disappear during the white dwarf eclipse; 
%At most about 1/3 of
%the blue absorption wing of \Hbeta\ can come from the disc. 
we
conclude that these features come predominantly from the white dwarf
\nocite{MarshEt:1987}(c.f.  Marsh {et~al.} 1987). 
The absorption cores remain during the white dwarf eclipse,
indicating they arise at least partially in the disc.

%{\sout{The roughly V-shaped white dwarf
%absorption {\color{red}will} also contribute to the deep cores in the
%Balmer lines, but we note that \nocite{MarshEt:1987}Marsh {et~al.} (1987) believes any low
%velocity absorption line components in Z\,Cha must come from the outer
%disc, along the line of sight to the white dwarf.}}

%{\color{red}\sout{Estimating the systemic velocity from H$\alpha$ in the March 2000
%spectra by fitting a double-Gaussian profile to the double peaks where
%the line is least affected by the hotspot emission
%\nocite{RolfeThesis:2001}(Rolfe 2001) gives $\gamma=13.6$\,\kms. The result is
%highly dependent on the phase range used, the resolution of the
%observations is only $\sim$200--300\,\kms\ and radial velocity
%standards were not used. \nocite{WuEt:2001}Wu {et~al.} (2001) estimated
%$\gamma=-4\pm32$\,\kms. We therefore adopted a value $\gamma=0$\,\kms\ 
%for the analysis in this paper.}}

\subsubsection{The hotspot spectrum}
\label{HotSpotSpectrum}

\begin{figure}
  \centerline{\epsfig{file=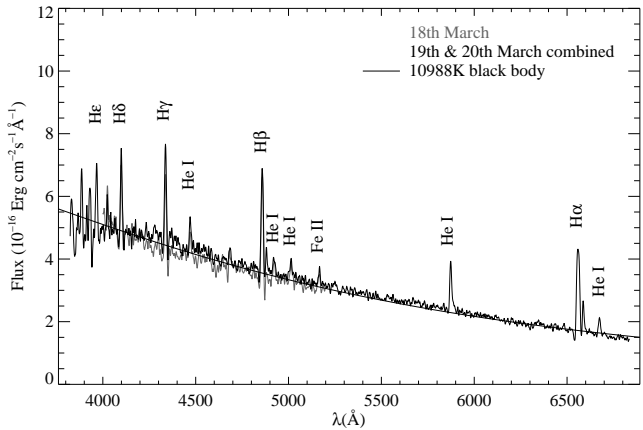,width=\columnwidth}}
  \caption{
    The spectrum of the orbital hump, obtained by subtracting the
    average spectrum at hump minimum from the average at hump maximum
    (smoothed as in Fig.
    \ref{AvSpectraFig2000}).  The smooth curve
    is the blackbody fit.}
  \label{HotSpotSpectrumFig}
\end{figure}

The very high inclination of IY\,UMa 
\nocite{SteeghsEt:2003}(c.f.  {Steeghs} {et~al.} 2003)
leads to a prominent orbital hump
as the hotspot comes into view. The hotspot spectra in Fig.
\ref{HotSpotSpectrumFig} 
%{\sout{(one for 18 March 2000 and one for 19 and 20
%March combined)}} 
were produced by subtracting the average spectrum
around hump minimum (orbital phases 0.5 to 0.6) from that around hump
maximum (phases 0.75 to 0.9).
Strong emission lines in the Balmer series and weaker emission in
\HeI\ and \FeII\ are superimposed on a blue continuum. The lines show
one strong peak indicating a concentrated emission region, with some
weaker
peaks and dips 
%{\sout{in some due to the}
arising from changes in 
%{\color{red}
line profile
  between the hump maximum and minimum. 
%{\sout{due to intrinsic temporal
%  changes in the emission and to our changing view of the line
%  emitting regions throughout the orbit.}} 
%\sout{The quiescent spectra of
%Z\,Cha, OY\,Car and V2051\,Oph showed no line emission from the
%hotspot \nocite{MarshEt:1987,BaileyEt:1981,WattsEt:1986}(Marsh {et~al.} 1987;  ,) although a
%number of other systems have done e.g. WZ\,Sge, U\,Gem, V893\,Sco and
%IP\,Peg
%\nocite{KrzeminskiEt:1964,SpruitEt:1998,Smak:1976,MatsumotoEt:2000,MarshHorne:1990,WolfEt:1998}( ; Spruit \& Rutten 1998;  ,; Marsh \& Horne 1990;  ).}}

Fig~\ref{HotSpotSpectrumFig} shows a simple blackbody
  spectrum fit (assuming $d=190\pm60$\,pc, from P2000) which has a
  temperature $T$=10990$\pm$120\,K and emitting area
  $A=3.8^{+3.0}_{-2.1}\,\times 10^{18}\rm{cm}^2$
  ($=0.0015^{+0.0011}_{-0.0008}\,a^2$ where $a$ is the orbital
  separation). The emission lines were masked out. The error in $T$ is
  a 90 per cent confidence estimate where the confidence region was
  determined using a Monte-Carlo ``bootstrap''-type resampling and
  fitting of the data 30000 times. The error in $A$ is dominated by
  propagating the uncertainty in $d$. Systematic errors are not
  accounted for in these confidence limits, e.g. errors in the
  atmospheric extinction correction and any deviation of the real
  spectrum from the assumed blackbody form. The latter is hard to
  detect since we only sample the tail of the Rayleigh-Jeans
  distribution.  \nocite{MarshEt:1987}Marsh {et~al.} (1987) produced a hotspot spectrum for
Z\,Cha using the same technique, finding a non-blackbody continuum,
with a significant drop below 4000\AA, probably due to Balmer
absorption. 
%{\sout{Our spectra do not cover this range.}} 
The effective area
determined here is consistent with the simple stream-disc impact model
used for IY\,UMa in \nocite{RolfeEt:2001a}Rolfe {et~al.} (2001b). The temperature
%  {\sout{found is typical for such systems which are typically}}
is typical: hotspot temperatures of 
  $\sim$10000--20000\,K
were found by
  \nocite{StanishevEt:2001,RobinsonEt:1995,WoodEt:1986}Stanishev {et~al.} (2001); Robinson {et~al.} (1995); Wood {et~al.} (1986).

\subsection{2001 January: IY\,UMa rises to normal outburst}
\label{AvSpec2001}

\begin{figure}
\centerline{\epsfig{file=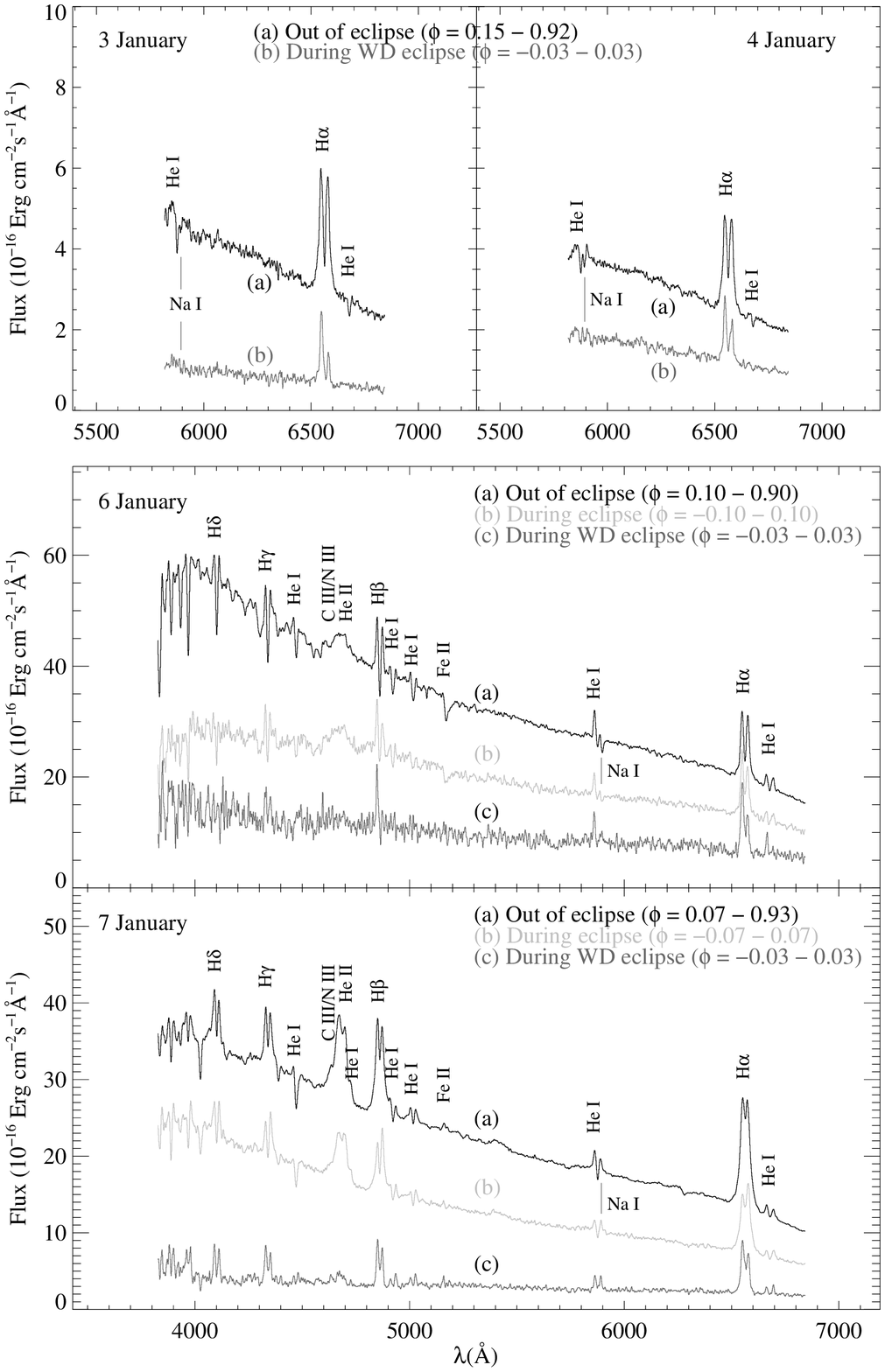,width=\columnwidth}}
\caption{
  Average spectra of IY\,UMa during quiescence, rise, and normal
  outburst in Jan 2001.}
\label{AvSpectraFig2001}
\end{figure}

Fig. \ref{AvSpectraFig2001} shows the average spectra;
%{\sout{ from the Jan 2001
%run.} 
for 
2001 Jan 3 and 4 we show just the blue end of the red spectra
which were discussed in \nocite{RolfeEt:2002a}Rolfe {et~al.} (2002b). 
The phase ranges used are given in Fig.~\ref{AvSpectraFig2001}.
%and are grouped,
%averaged, and smoothed as those from March 2000. The 6 and 7 Jan
%spectra are grouped, averaged and smoothed into eclipse and
%non-eclipse phase ranges.
%
On 2001 Jan 3 and 4 the system is quiescent, the only notable
difference from 2000 March being the deeper core between the disc
peaks.

On 2001  Jan 6 the system is 7-8 times brighter. Deep core absorption
is seen in all emission lines except \Halpha.
%{\color{red}\sout{(where
%    the core is nevertheless much deeper than in quiescence), and
%    completely dominates the emission in \HeI\ (except \HeI\ at
%    5876\AA) and \FeII.}} 
This resembles OY\,Car during a normal
outburst \nocite{HarlaftisMarsh:1996b}(Harlaftis \& Marsh 1996) and some superoutburst spectra
of Z\,Cha \nocite{HoneyEt:1988}(Honey {et~al.} 1988). The Balmer absorption
  wings are no longer seen, meaning that the deep core absorption in
  outburst cannot be due to the white dwarf. This also adds weight to the
  suggestion that the cores in quiescence are not entirely due to the
  white dwarf absorption. The 4640\AA\ \CIII/\NIII\ Bowen blend and
\HeII\ 4686\AA\ have appeared, as in IY\,UMa in
superoutburst \nocite{WuEt:2001}(Wu {et~al.} 2001), in other dwarf novae in outburst
\nocite{MoralesRuedaMarsh:2002}(Morales-Rueda \& Marsh 2002) and in nova-likes e.g. V348\,Pup and
UX\,UMa \nocite{TuohyEt:1990,RolfeThesis:2001,RuttenEt:1993}(Tuohy {et~al.} 1990; Rolfe 2001; Rutten {et~al.} 1993). The
\HeII\ emission probably results from the reprocessing of EUV and
X-ray boundary layer emission \nocite{PattersonEt:1985}(Patterson \& Raymond 1985), while the
\NIII\ emission may come from conversion of \HeII\ \Lyalpha\ 
transition photons to \NIII\ photons via \OIII\ 
\nocite{Deguchi:1985}(Deguchi 1985).
%{\color{red}\sout{The line profile is broad like
%    the Balmer lines, suggesting a source in the disc, although the
%    breadth could result from the blending.}}

On 2001 Jan 7 the continuum is about 30 per cent less than Jan 6 and
the 
line absorption cores 
%{\color{red} 
%{\sout{in the emission lines}} 
are much weaker.
%\sout{,
%    particularly in \Halpha\ and \Hbeta, and has disappeared in \FeII.
%    Core absorption remains only in \HeI.}} 
\HeII\ 4686\AA\ and the
Bowen blend are much stronger, with \HeII\ comparable in strength to
\Hbeta. \HeII\ 5412\AA\ is faintly visible. The \HeII\ 4686\AA\ line
profile is consistent with strong double-peaked emission from the disc
blended with a \CIII/\NIII\ and \HeI\ 4713\AA\ contribution. The most
obvious absorption feature is \HeI\ 4471\AA.  This has low velocity,
is very narrow, and disappears almost precisely during the WD
eclipses (Fig. 4), implying 
%{\sout{that this is}} 
absorption 
occuring close to the WD. 
%{\sout{or very close to it.}  
%{\color{red}\sout{It is not possible to come to
%    such a conclusion for this line on 6 Jan due to contamination from
%    the emission component and lower S/N.}}

%Clearly, 
IY\,UMa went into outburst between Jan 4 and Jan 6, and
continued in outburst on Jan 7. The accretion rate onto the white
dwarf must have been several times higher on Jan 7 than on Jan 6,
increasing the flux of X-rays and EUV emitted from the boundary layer,
thus enhancing the strength of the \HeII\ emission.  
Meanwhile 
%{\sout{Note}} 
the optical
continuum 
%{\sout{was already}} 
decreased between 2001 Jan 6 and 7, 
indicating
the outer, optical continuum emitting disc was declining from its
peak mass accretion rate. This indicates an outside-in outburst.
IY\,UMa 
%{\sout{had}} 
returned to quiescence by Jan 9 or 10 (D.
Steeghs, private communication), suggesting this was a normal
outburst.

\subsection{2000 January: IY\,UMa in superoutburst}

\begin{figure}
\centerline{\epsfig{file=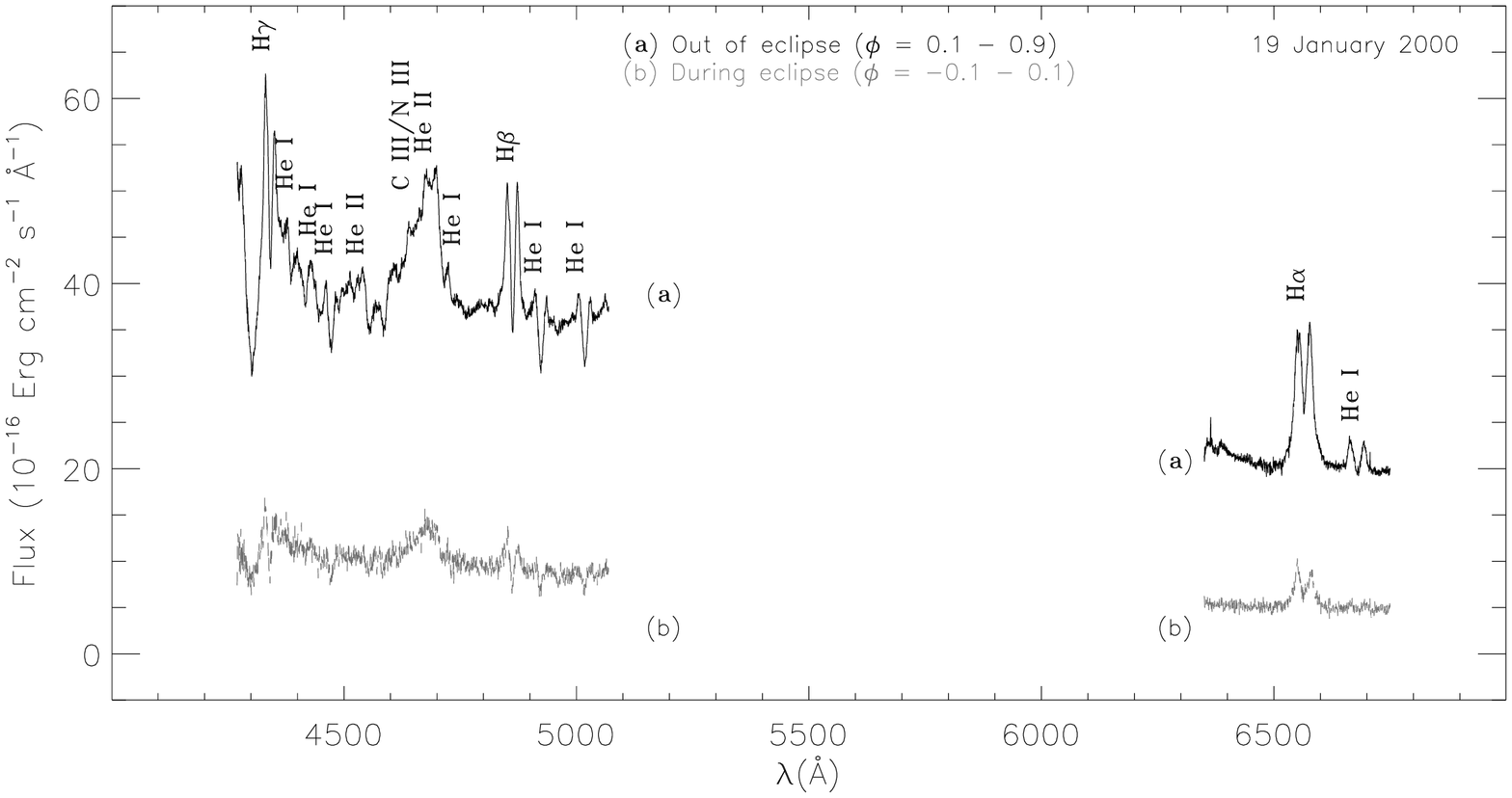,height=\columnwidth,angle=90}}
\caption{Average spectra of IY\,UMa during superoutburst.}
\label{lmr:avsp}
\end{figure}

The average red and blue spectra 
%{\sout{of IY\,UMa during superoutburst are presented}} 
in Fig. \ref{lmr:avsp} show a variety of hydrogen and helium
lines 
%{\sout{are}} 
superimposed on a blue continuum. The lines appear
double-peaked with very deep core absorption:
%{\sout{in the core of the line, the}}
absorption dominates emission in all \HeI\ lines 
%{\sout{apart from}}
except \HeI\ 
6678\AA. The most striking feature is the emission in \HeII\ and the
Bowen blend. 
%{\sout{which is of similar strength to \Hbeta, as on 7 Jan
%2001. There is also another interesting feature to the blue}
Bluewards of
\Hgamma\  is a strong absorption that was also hinted in the 2001 Jan 6
normal outburst spectra. 
%{\sout{but that is even deeper in this case.}} 
No such
feature was visible on 2001 Jan 7. The spectrum looks very similar to
that during normal outburst on the 2001 Jan 6, apart from the stronger
\HeII/Bowen emission.

\subsection{Equivalent Widths}

Table \ref{ew} gives emission line equivalent widths (EWs) 
%{\sout{measured for the
%emission lines}} 
during the two outburst states. These values 
%{\sout{ have been
%measured}} 
are from the average out-of-eclipse spectra for each night shown
in Figs. \ref{AvSpectraFig2001} and \ref{lmr:avsp}. 
%{\sout{Positive values
%correspond to net emission in the line.}}

During the 2001 Jan normal outburst the EWs of \Halpha, \Hbeta\ and
the \HeII/Bowen/\HeI\ blend increase considerably from Jan 6 to Jan 7, 
by factors $\sim$2--4. The EW of \Hgamma\ is negative on Jan 6 due to
the 
%{\sout{interference of the}} 
deep, blue absorption feature. \HeI\ 5016\AA\ 
is similarly diminished by core absorption on Jan 6. \HeI\ 6678\AA\ 
barely changes in EW during the Jan 2001 observations. During the Jan
2000 superoutburst \Halpha\ has a similar EW to that on 2001 Jan 6.

The observations were taken at different stages in two different types
of outburst, with the rapidly changing accretion flow leading to
complicated changes in the continuum, absorption and emission lines.
If irradiation powers the emission lines (c.f. section 3.2)
%as suggested in the previous section, 
while viscous dissipation produces the continuum,
%then we expect the emission line flux to be uncorrelated with the
%continuum and 
the EW 
%{\sout{to potentially change}}
could vary considerably. 
%{\sout{throughout outburst.}} 
This is indeed the case for the Balmer lines and \HeII/Bowen
feature, while the absorption dominated \HeI\ lines are inconclusive.
%{\sout{The positive and negative contributions to the equivalent widths of
%\HeI\ lines are of comparable size, hence the net value is noisy.
%Clearly the absorption line flux should always be correlated with the
%continuum.}}

\begin{table}
\caption{Equivalent widths (in \AA) of the lines present in the
  spectra during the two outburst states studied.}
\label{ew}
\begin{center}
\begin{tabular}{l|c|c|c}
\hline
& \multicolumn{2}{c}{Outburst} & Superoutburst \\
\hline
                                  & 06/01/01         & 07/01/01         & 19/01/00\\
\hline
\Halpha                           &  23.08$\pm$0.14  & 56.30$\pm$0.11   & 28.54$\pm$0.12  \\
\Hbeta                            &   4.83$\pm$0.12  & 18.19$\pm$0.08   &  9.29$\pm$0.09  \\
\Hgamma                           &  -0.95$\pm$0.12  &  8.77$\pm$0.09   & 11.66$\pm$0.12  \\
    He\,{\sc i}\,$\lambda$5016    &   0.07$\pm$0.09  &  1.65$\pm$0.06   &  0.72$\pm$0.07  \\
    He\,{\sc i}\,$\lambda$6678    &   2.72$\pm$0.11  &  2.69$\pm$0.08   &  3.98$\pm$0.09  \\
\hspace{-0.7cm}\begin{tabular}{l}
He\,{\sc ii} 4686\AA\ \\
+He\,{\sc i}\,$\lambda$4713 \\
+Bowen blend\\
\end{tabular}$\Bigg\}$&   4.82$\pm$0.16  & 22.50$\pm$0.11   & 25.98$\pm$0.15  \\
\hline
\end{tabular}
\end{center}
\end{table}

\section{Lightcurves}
\label{Lightcurves}

\begin{figure*}
\centerline{\epsfig{file=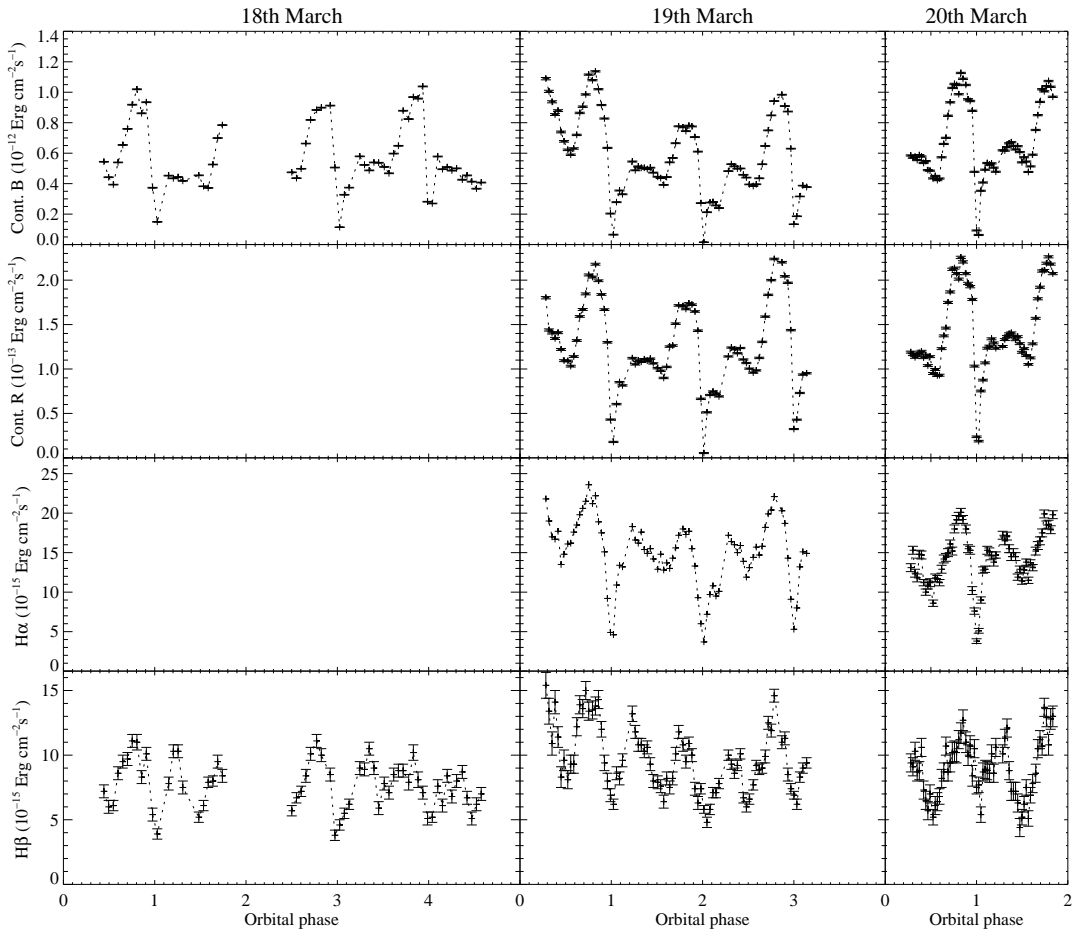,width=0.8\textwidth}}
\caption{March 2000 quiescence average orbital lightcurves}
\label{LightcurvesFig2000}
\end{figure*}

Lightcurves of the continuum and continuum-subtracted emission lines
are shown in Fig. \ref{LightcurvesFig2000} (2000 March) and Fig.
\ref{LightcurvesFig2001} (2001 Jan). ``Continuum B'' is the flux
integrated in the range 4000\AA--5200\AA\ (covered by all the March
2000 spectra) while ``Continuum R'' is over the range
5950\AA--6450\AA\ (covered by all the 2000 and 2001 spectra except 2000
March 18). For \Hbeta, the average flux either side of the profile
was subtracted to reduce the effect of the absorption wings. Note that
slit loss corrections were not possible for the 2000 March 
observations, but the data suggest that apart from the omitted phase
range 2--2.5 on 18 March, conditions were photometric. The 2000 January
observations conditions were not photometric and no slit loss
correction was possible, so we do not present lightcurves for these
data.

\subsection{2000 March: quiescence}
\label{Lightcurves2000}

The continuum lightcurves for 2000 March reveal the strong orbital
hump and eclipse seen in photometry
\nocite{RolfeEt:2001a}(P2000, Rolfe {et~al.} 2001b). On 2000 March 19 and 20 there is a
double-humped structure, with the usual strong orbital hump peaking
around phase 0.8--0.9 and a weaker hump peaking around 0.3--0.4.  The
strong hump is the stream-disc impact coming into view on the near
side of the disc while the weak hump is when the stream-disc impact is
again seen roughly length ways, but on the far side of the disc. 
%{\sout{Such double-humped}}
Similar orbital curves have been seen in WZ\,Sge and
AL\,Com \nocite{RobinsonEt:1978,PattersonEt:1996}(Robinson, Nather \&  Patterson 1978; Patterson {et~al.} 1996), but with both humps
having the same amplitude. \Halpha\ and \Hbeta\ 
%{\sout{curves are similar, but with}} 
show a stronger secondary hump and a shallower eclipse,
the latter resulting from disappearance of the
unsubtracted core of the white dwarf Balmer absorption lines
during eclipse (an effect which is stronger for \Hbeta).

\subsection{2001 January: normal outburst}
\label{Lightcurves2001}

\begin{figure*}

  \centerline{\epsfig{file=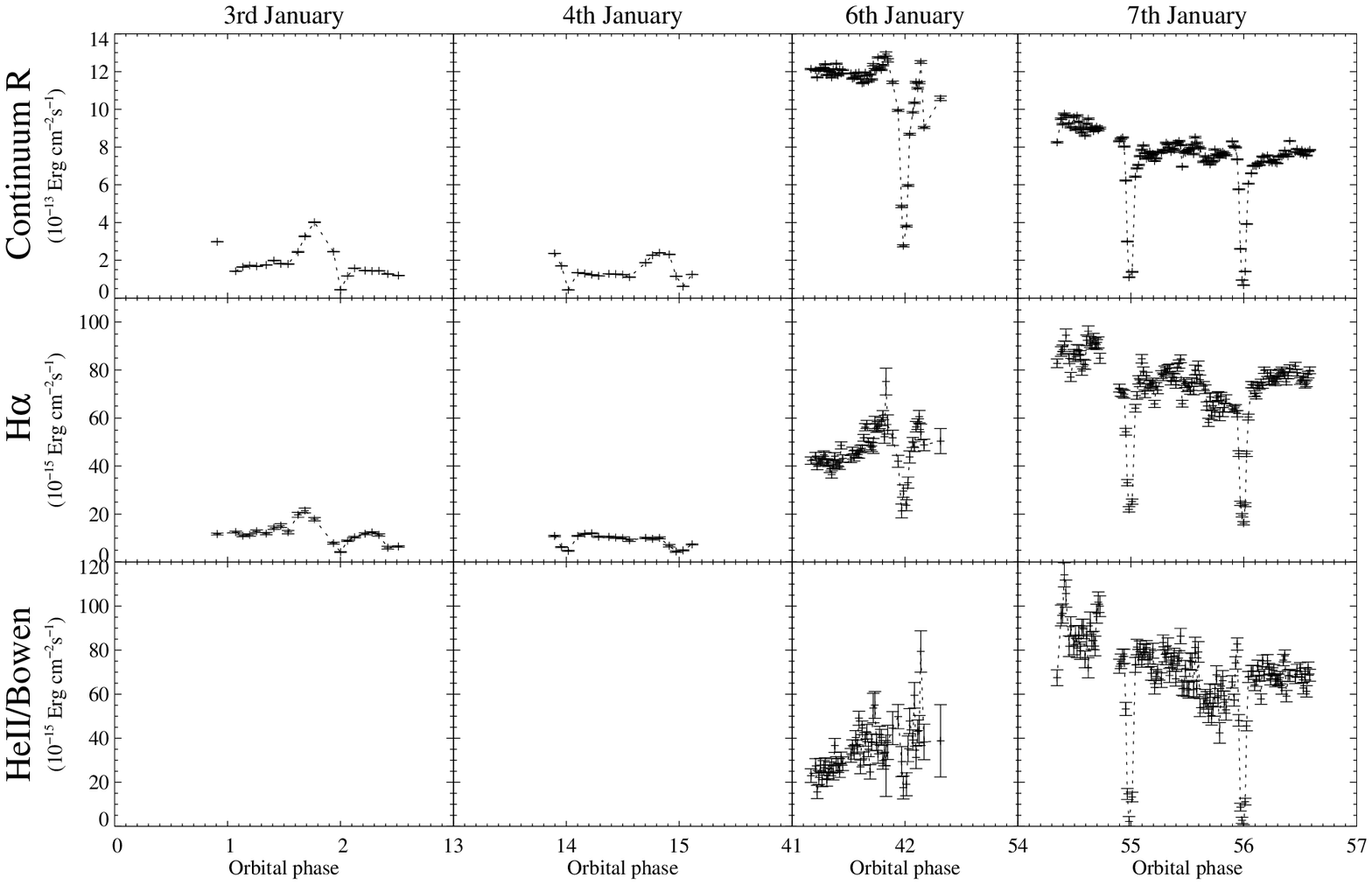,width=0.9\textwidth}}
\caption{Jan 2001 lightcurves during quiescence, rise, and normal
  outburst.}
\label{LightcurvesFig2001}
\end{figure*}

The Jan 3 and 4 continuum curves (Fig. 7) show 
%{\sout{just the usual}}
a single orbital
hump;
%{\sout{ with the}
the flux away from hump and eclipse is within a factor
$\sim$2 of March 2000. The hump is about twice as bright on Jan 3 as on
Jan 4. The \Halpha\ curves 
%{\sout{are more unusual, with the orbital hump and
% eclipse clear} 
on Jan 4 shows no sign of the hump.
%{\sout{ on the 4th.}}
Though low
time resolution and phase coverage prevents detailed analysis,
%{\sout{ of these lightcurves, but they do suggest}} 
we see
changes in the behaviour of the
system over one day.  
If it is linked to
the following outburst,
this is important. 
No advanced warning of an impending outburst
has been seen before.

%Debate about the mechanism(s) driving dwarf nova outbursts has
%continued for over two decades.  The accretion disc thermal viscous
%instability model is regarded as being responsible for the outbursts,
%and yields comprehensive quantitative explanations of the observed
%characteristics of outbursts in dwarf novae and in their neutron star
%and black hole analogues (see e.g. \nocite{HynesEt:2002} ()).  Some authors
%augment this disc instability with an enhanced mass transfer rate from
%the mass donor star in outburst, a modification which excites vigorous
%debate, e.g. \nocite{OsakiMeyer:2003}Osaki \& Meyer (2003). A change in mass transfer rate
%can be inferred from a change in the continuum emission from the
%stream-disc impact region if one assumes the geometry does not change
%(c.f. \nocite{FoulkesEt:2004}Foulkes {et~al.} (2004)). Surprisingly this simplistic
%interpretation of our R band continuum light curves suggest the mass
%transfer rate on 3rd Jan 2001 is comparable to that in quiescence
%(i.e.  March 2000, as indicated in Fig. 6) and {\bf decreases}
%immediately before the rise to outburst.

Between Jan 4 and 6, the continuum flux rises by a factor of $\sim$7,
decreasing again by the beginning of the Jan 7's observations.
During Jan 7 the flux drops by a further 20
percent. The eclipses are all deep, with the eclipse during orbit 42
broader and shallower than those during orbits 55 and 56.
This implies the light distribution becomes more concentrated
around the white dwarf between Jan 6 and 7;  suggesting
the outer disc 
%{\sout{regions are now}} 
is returning to  quiescence. 
Measurement by eye yields full eclipse widths which correspond to
emission distributions of radius 0.37$\pm$0.10$a$ on Jan 6 shrinking to
0.21$\pm$0.05$a$ on Jan 7. These radii are both much smaller than the 3:1
resonance radius (0.46$a$) indicating a normal outburst.
%{\sout{ which must be populated in a superoutburst}
n.b. P2000 measured radius 0.44$\pm$0.03$a$ in superoutburst. 
An unrealistic eccentricity ($\sim0.8$)
is needed for a non-circular disc to explain the change in
eclipse width.
%{\sout{ The
%lightcurves of the emission lines are very similar to one another,
%although they are noisier for \Hbeta\ and the \HeII/Bowen blend than
%for \Halpha.}}

The 
%{\sout{continuum and emission line}} 
eclipses are
%{\sout{appear}} 
fairly symmetrical,
and centred on phase 0, suggesting emission centred on the white
dwarf.  In contrast with the continuum, the emission lines 
%{\sout{are still}}
all rise on Jan 6, being nearly twice as bright at the start of the
observations on Jan 7 than during Jan 6. These changes are not
simply a result of variations in the core absorption. 
%{\sout{Our observations show that}} 
The delay in rise of the lines relative to the continuum is
between about one orbit and two days. On the 2001 Jan 7 the 
%{\sout{fluxes decrease more slowly than the continuum, their equivalent widths}}
EWs in
\Halpha\ and \HeII\ increase by about 7--8 per cent. The \Halpha\ 
eclipses suggest the same decrease in radius of the line emission
region as for the continuum. The \Halpha\ EW doubles
during eclipse, while for \HeII\ it decreases almost to zero. 
%{\sout{as
%expected from the lightcurves, where the \Halpha\ eclipse is shallower
%than that of the continuum while the \HeII\ eclipse is deeper. This
%tells us the}} 
We conclude 
the \HeII\ emission comes from a smaller area than
\Halpha, concentrated closer to the white dwarf.

\section{Doppler tomography}
\label{Tomography}

Doppler tomography \nocite{MarshHorne:1988}(Marsh \& Horne 1988) exploits phase-resolved
emission line profiles to produce velocity-space maps of the emission
distribution. It assumes the emitting material and its velocities are
in the orbital plane, that the brightness and visibility of each point
do not vary with phase, and that the intrinsic line profile is narrow
compared with the Doppler shifts due to the velocity distribution.
We excluded data from phases -0.1 to 0.1 
%{\sout{were left out}} 
since eclipses occur then.
The hotspot also has a varying brightness, so Doppler maps indicate
only an orbital average of the hotspot emission. The Fourier-filtered
back projection technique was used to produce the maps.
Levels below the continuum are shown the same as the continuum level.
%We discuss each distinct group of observations in turn.

\subsection{2000 March: quiescence}
\label{Tomography2000}

\begin{figure*}
\epsfig{file=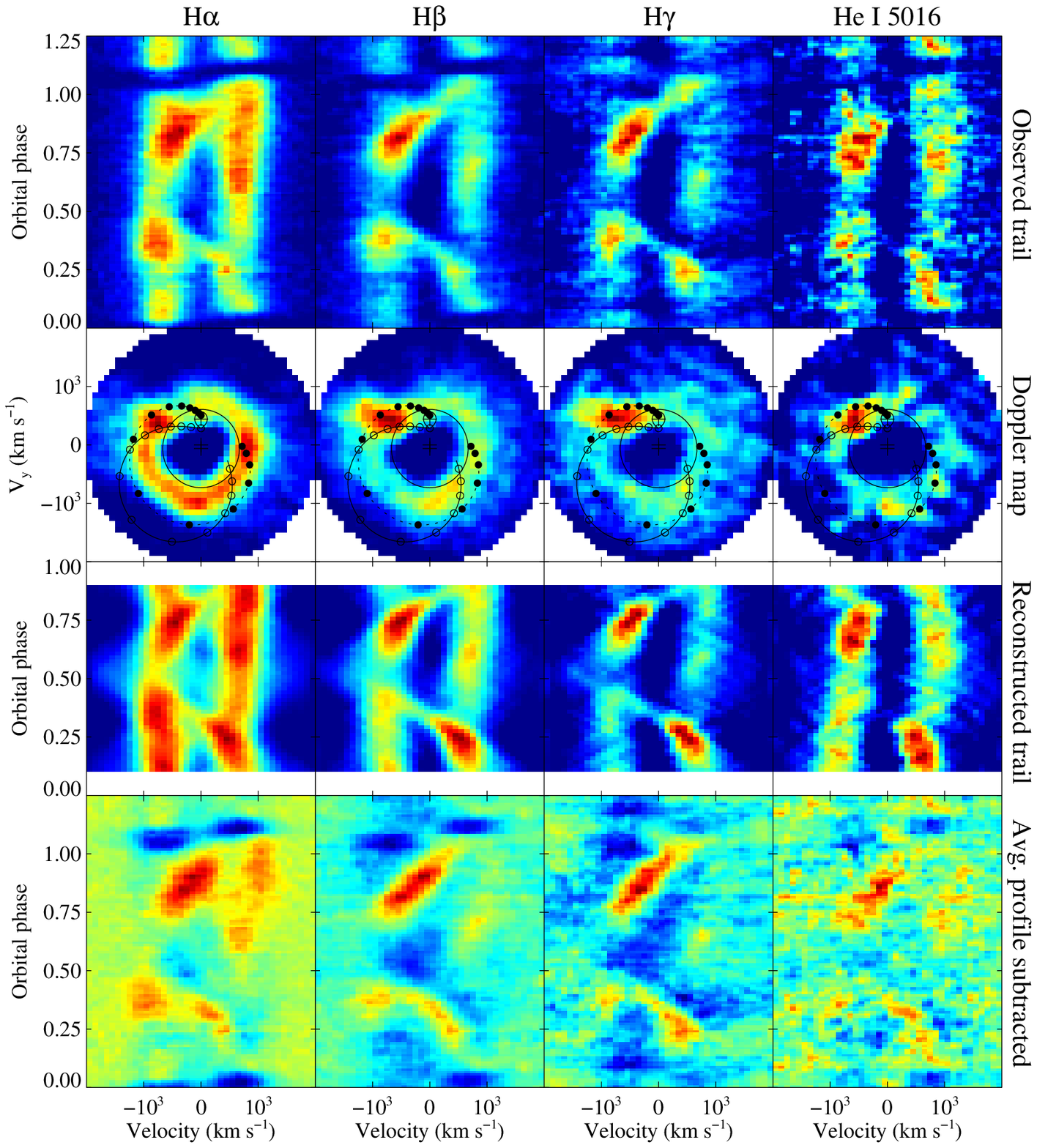,width=0.8\textwidth}
\caption{
  Doppler tomography combining all 3 nights of data during quiescence
  in March 2000. \emph{Top row:} Phase-folded, phase-binned,
  velocity-binned and continuum-subtracted trailed spectra. Phases
  0--0.25 are repeated. \emph{Second row:}
  Fourier-filtered back-projections. Emission from the donor star
  should appear within the black teardrop. Disc emission should appear
  outside the black circle (the Keplerian velocity at the radius where
  the disc is tidally truncated).  The ballistic stream velocity is
  the arc with unfilled circles, and the arc with filled circles is
  the Keplerian velocity along the stream trajectory. These markings
  assume the orbital parameters of P2000.  \emph{Third row:} Trailed
  spectra reconstructed from Doppler maps. \emph{Bottom row:} Observed
  trail minus average line profile in phase range 0.1 to 0.9.}
\label{Tomography2000Fig}
\end{figure*}

The trailed spectra in 2000 March revealed the same orbital modulation
throughout. We therefore consider the phase-folded trails
shown in the top row of Fig. \ref{Tomography2000Fig}.  Double-peaked
emission with eclipse first in the blue peak and then in the red, as
expected from a prograde disc, is seen in the Balmer
lines. 
%{\color{red}\sout{In the \HeI\ line we see the blue peak
%    (negative velocity), the red peak being mostly cancelled by the
%    \NaI\ absorption.}} {\color{red}\sout{In all four lines, }}
The disc emission in the
red wing around phase 0.3 is weaker than at other phases (except
eclipse), with this being very clear in \Hgamma. A similar effect was
seen in WZ\,Sge by \nocite{SpruitEt:1998}Spruit \& Rutten (1998), where an S-wave of
absorption was seen delayed in phase relative to the hotspot S-wave.
The brightening in the blue wing around phase 0.3 is a distinct
feature, not simply the coincidence of the stream and disc emission.
This could result from a bright region of the disc on the opposite
side from the donor.  Between the disc peaks is the core and white
dwarf Balmer absorption.  A strong S-wave with phase-dependent
brightness and the same phasing and amplitude appears in all four
lines. The strength of the disc emission is similar to that of the
S-wave emission in \Halpha, but the relative strength of disc to
S-wave decreases for higher order Balmer lines, with the brightest
parts of the S-wave dominating 
%{\sout{the disc emission}}
 in \Hgamma.
%{\color{red}\sout{The S-wave is also dominant in \HeI. Its variation
%    in brightness is clearest in the \HeI\ line and} 
The bottom panel
  of Fig. \ref{Tomography2000Fig} shows the trails after subtracting the
average
  line profile in the phase range 0.1--0.9. This removes
  much of the disc profile and core absorption, revealing that S-wave
  brightness
%} 
follows that of the orbital hump (c.f. Fig. 6). 
The S-wave is eclipsed
late indicating an origin at the stream-disc impact. The sharp
brightening in the S-wave/red wing around phase 0.2 results from poor
phase coverage around phase 0.20-0.25 on 19 and 20 March.

%{\sout{The second row of Fig. \ref{Tomography2000Fig} shows the Doppler maps
%from the March 2000 data. All four maps}}
The four Doppler maps (Fig.~\ref{Tomography2000Fig}, Row 2) 
show absorption at low
velocity corresponding to the deep cores visible in the average
spectra. The \Halpha\ map shows emission from disc material within the
tidal radius. The stream-disc impact is seen between the two arcs
corresponding to the stream trajectory and its Keplerian velocity
(c.f. U\,Gem, \nocite{MarshEt:1990b}Marsh {et~al.} (1990) and WZ\,Sge, \nocite{SpruitEt:1998}Spruit \& Rutten (1998)).
This is also seen in SPH simulations where more detail is apparent
\nocite{FoulkesEt:2004}(Foulkes {et~al.} 2004). The \Halpha\ disc emission is weaker where the
two arcs cross it. \nocite{SpruitEt:1998}Spruit \& Rutten (1998) point out that
%{\color{red}\sout{in WZ\,Sge}} 
the stream-disc impact is hot enough to
ionize hydrogen, suppressing Balmer emission until further downstream
where the hydrogen has recombined. If the hydrogen in the converging
stream and disc flows is ionized, then we expect the Balmer emission
from this region to be suppressed. The light curves of
\nocite{SteeghsEt:2003}{Steeghs} {et~al.} (2003) suggest that IY UMa's extreme edge-on
inclination affords us a view into a hot shock-heated cavity driven
into the disc by the stream impact.  SPH simulations
\nocite{FoulkesEt:2004}(Foulkes {et~al.} 2004) suggest that the dissipation at the impact
occurs in both the stream and the disc flows, and is localised at the
impact velocity of each. The March 2000 \Halpha\ map is suggestive of
ionisation caused in this way. The $\sim$11000\,K blackbody hotspot
temperature found in IY\,UMa (c.f. Fig. 3)
would be sufficient to ionize hydrogen.
%{\sout{The \Hbeta\ and \Hgamma\ maps again show the disc emission and
%hotspot, but in these maps the hotspot is much stronger relative to
%the disc than in \Halpha. The \HeI\ 5876\AA\ map shows just the very
%strong bright spot with the weak disc emission cancelled out by the
%\NaI\ absorption.}}{\color{red}\sout{ The Balmer maps resemble those of
%    WZ\,Sge \nocite{SkidmoreEt:2000a,SpruitEt:1998}( ; Spruit \& Rutten 1998).}}

The third row of Fig. \ref{Tomography2000Fig} shows the line profiles
reconstructed from the Doppler maps i.e. the line profiles we would see
if the emission distribution was exactly as shown in the Doppler map,
and all assumptions of Doppler tomography were valid. Comparing the
reconstructed trails with the observed ones provides a useful measure
of how reliable the Doppler maps are. Eclipses (not accounted for in
Doppler tomography) are omitted, but we see clearly in the
reconstructed trails the same disc and hotspot features as in the
observations, but with features resulting from intrinsic variations
during the orbit (the hotspot orbital hump) not correctly reproduced,
since the Doppler map is an orbital average. The Doppler map has
attempted to reproduce the orbital hump by superposing sine waves of
differing velocity amplitude. This leads to the emission extending out
to $\sim$2000\kms\ at orbital phase 0.5, for which there is no evidence
in the observed trails.

\subsection{2001 January: normal outburst}
\label{Tomography2001}

\begin{figure*}
  \epsfig{file=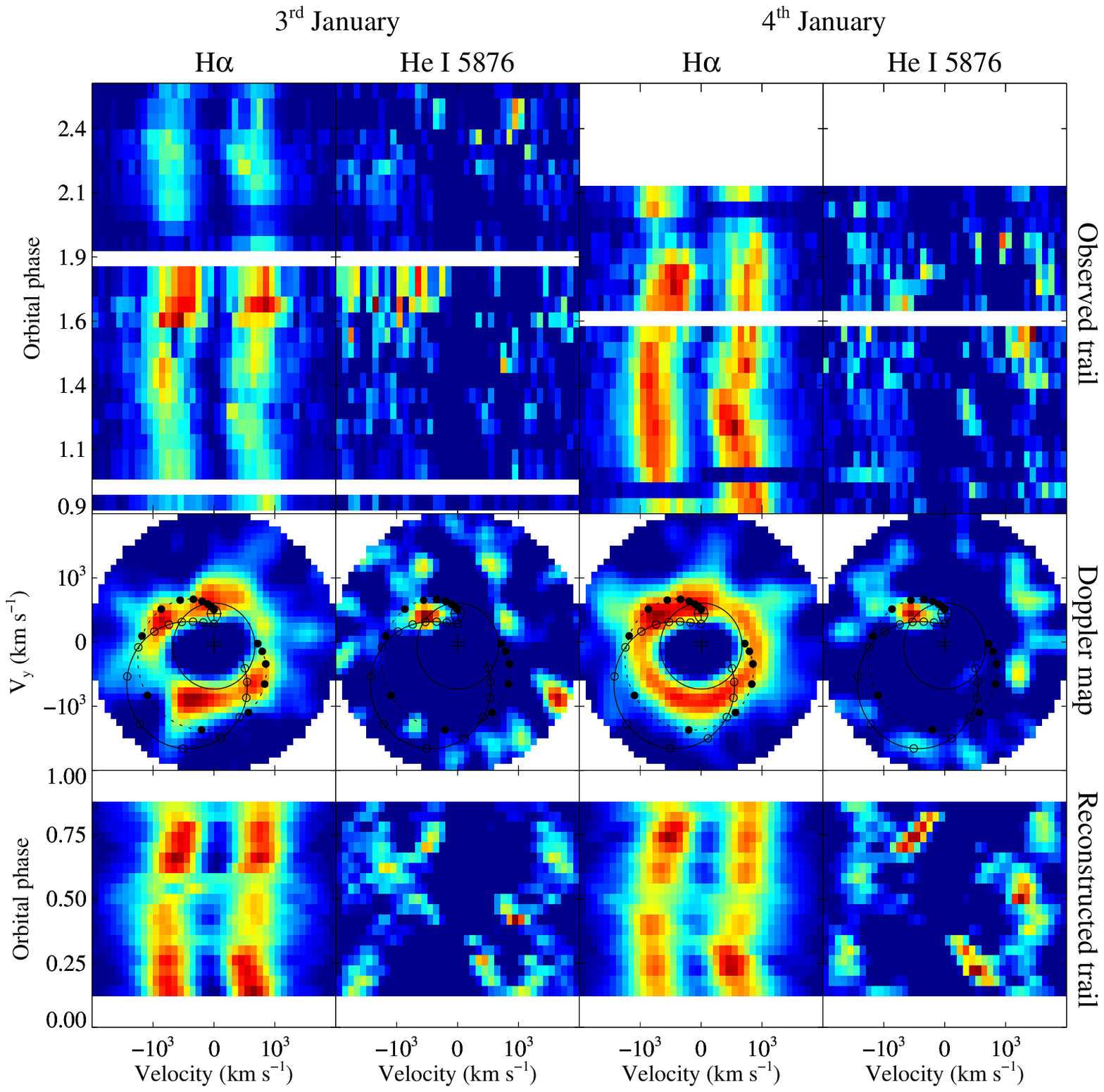,width=0.8\textwidth}
  \caption{
    Doppler tomography before outburst in Jan 2001. \emph{Top row:}
    Phase-binned, velocity-binned and continuum-subtracted trailed
    spectra. \emph{Other rows:} As in Fig. \ref{Tomography2000Fig}.}
  \label{Tomography34Fig}
\end{figure*}
In Jan 2001 we observed the accretion flow behaviour around the rise
from quiescence to normal outburst. 

\subsubsection{2001 Jan 3,4: pre-rise}
Trailed spectra from 2001 Jan 3 and 4 for \Halpha,
%\Hbeta, \Hgamma\ 
and \HeI\ 5876\AA\ are shown in Fig.
\ref{Tomography34Fig}. The wavelength of the weak \NaI\ absorption
doublet line at 5890--5896\AA\ corresponds to a velocity range of
about 700--1000\,\kms\ relative to \HeI\ 5876\AA\ so we must be wary
of this when studying 
%{\sout{the \HeI\ trails and Doppler maps.}} 
this line. 
%{\sout{On 3 and 4
%Jan} 
\Halpha\ shows double peaks and a blue-to-red eclipse; 
%{\sout{ of the disc.}} 
the S-wave 
%{\sout{from the hotspot}} 
is not visible, although brighter
regions in the peaks hint at its presence, particularly on 2001 Jan 4. 
%{\sout{The deep core and low S/N and time resolution prevents us from seeing
%detailed structure, particularly the S-wave crossing at low
%velocities}}. 
The low S/N in \HeI\ prevents any structure being seen in
that trail, apart from the core/\NaI\ absorption and possibly a bright
region around the blue peak between about phase 0.6 and 0.75, maybe
due to an S-wave.

Doppler maps for 2001 Jan 3 and 4
(Fig.  \ref{Tomography34Fig}) are blurred by $\sim20^\circ$ by the long exposures
and the limited number of spectra. 
%{\sout{The long exposure times of the 3 and 4 Jan spectra will lead to $\sim20^\circ$
%azimuthal blurring in the maps. Furthermore there are a limited number
%of spectra contributing to the maps (16 and 13 respectively), so we
%should not draw any conclusion about the fine detail in these maps.
%The \Halpha\ maps both show a ring of disc emission, with asymmetry in
%the 3 Jan map from the bright red and blue peaks around phase 0.7 in
%the line profiles. Both maps}} 
Despite this all 4 maps show evidence 
of the hotspot at the same
location as in 2000 March, 
%{\sout{. Despite the apparent lack of information in
%the \HeI\ trails, both the 3 and 4 Jan maps show a clear hotspot at
%the same location as March 2000 (}} 
with \HeI\ a little further upstream than the
\Halpha\ hotspot. 
%{\sout{). So, even with these low quality maps, the main
%features of the quiescent accretion flow are visible, as in March
%2000. Note that there was no orbital hump from the hotspot in the Jan
%4 emission line lightcurves (Fig. \ref{LightcurvesFig2001}).  The
%reconstructed trails are not inconsistent with those observed, and
%make clearer which features correspond to the hotspot S-wave, though
%again the high velocity blue emission artefact at phase 0.5 is seen.}

\subsubsection{2001 Jan 6: outburst}
\begin{figure*}
  \epsfig{file=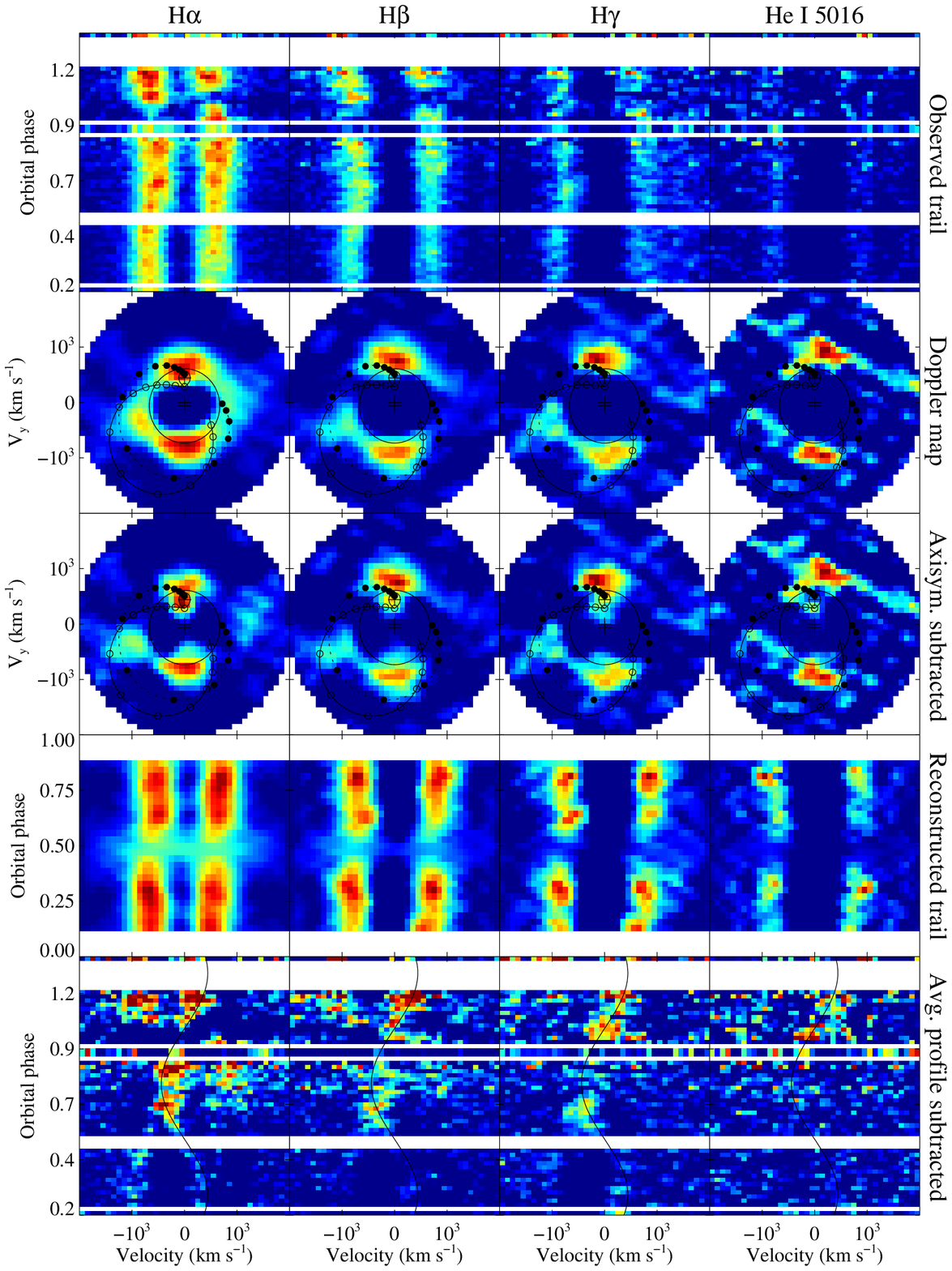,width=0.8\textwidth}
  \caption{
    Doppler tomography during rise/normal outburst on 6 Jan 2001.
    \emph{Top, second and fourth row:} As in Fig.
    \ref{Tomography34Fig}. \emph{Third row:} Non-axisymmetric
    component of Doppler maps in second row. \emph{Bottom row:} Observed trail minus average line profile in phase range 0.2 to 0.8, with donor velocity overplotted.}
  \label{Tomography6Fig}
\end{figure*}

The 2001 Jan 6 \Halpha, \Hbeta, \Hgamma\ and \HeI\ 5016\AA\ trails (Fig.
\ref{Tomography6Fig}) all show double-peaked disc emission, while
the blue-to-red eclipse is seen clearly in \Halpha\ and \Hbeta\ (the
strongest lines). 
%{\color{red}\sout{The red peak in the \HeI\ is
%    affected by the \NaI\ absorption.}} 
Apart from the rise in
brightness throughout the night, there is no other easily identifiable
behaviour.  
%{\sout{The very}} 
Deep absorption cores 
%{\sout{in the lines again}}
 hide any
structure between the peaks.

The 2001 Jan 6 Doppler maps (Fig. \ref{Tomography6Fig}, 2nd
row) differ considerably from those for 2001 Jan 3 and 4.  The
\Halpha\ map shows a ring of disc emission which has two bright
regions on opposite sides of the disc around maximum and minimum
$V_y$. The same structure is seen in \Hbeta\ and \Hgamma. The \HeI\ 
map is noisier but consistent. 
%{\color{red}\sout{but it also shows a
%    bright blob of emission in the top half of the lower left quadrant
%    of the disc; this extra emission is identifiable but weaker in the
%    Balmer maps.}} {\sout{Detailed consideration reveals} 
The two-fold
asymmetry is an artefact from leaving out the eclipse phases:
omitting the disc emission during eclipse weakens the left and right
sides of the ring in the map, while missing the very strong core
absorption during eclipse enhances the strip around $V_x=0$ in the
map, accentuating the top and bottom of the disc ring.  
The brightening of the line flux during the night (c.f. Fig.~7)
biases the map, and may thus contribute to the two-fold asymmetry.
We therefore made a set of maps (Rolfe, 2001) using spectra 
normalised so the total line flux remained constant with time. 
\Halpha\ showed similar asymmetries to those of the maps in
Fig.~\ref{Tomography6Fig} while the other lines produced inferior maps
due to the effects of noise in the intergrated line flux.
Accordingly we conclude the maps from 2001 Jan 6 show no more than emission 
from a disc within the tidal radius.

Each map for 2001 Jan 6 had its azimuthally averaged map 
%{\sout{(centred on the white dwarf velocity)}} 
subtracted
(Fig. \ref{Tomography6Fig}, third row), 
%{\sout{The resulting maps, shown in the
%, thus have features
%axisymmetric about the white dwarf removed,} 
enhancing non-axisymmetric
features.  This removes much of the disc ring and the central
absorption. The Balmer maps clearly show emission coming from the
velocity of the donor star. The quality of the \HeI\ map is too poor
to identify or rule out such a feature. The reconstructed trails show
clearly the disc emission and core absorption, with the double peaks
disappearing around phase 0.5 in contradiction with the observations,
%a result of omitting the eclipse data, 
confirming the conclusion that
the asymmetry in the Doppler maps is an artefact of omitting eclipses.

We subtracted the average line profiles for the phase
  range 0.2--0.8 from the observed trails 
(Fig.  \ref{Tomography6Fig}, bottom row),
reducing the disc and absorption
  components.
Overplotted is the expected velocity of the
  donor star. There is clear evidence of a weak emission component in
  \Halpha\ following the donor velocity. 
%{\sout{but nothing so clear in the
%  other, noiser, trails.}} 
This confirms that there is emission from the
  donor velocity, and that this is not merely an artefact in the Doppler
  maps, at least in the case of \Halpha.

\subsubsection{2001 Jan 7: Decline}
\begin{figure*}
  \hspace*{-1.5cm}\epsfig{file=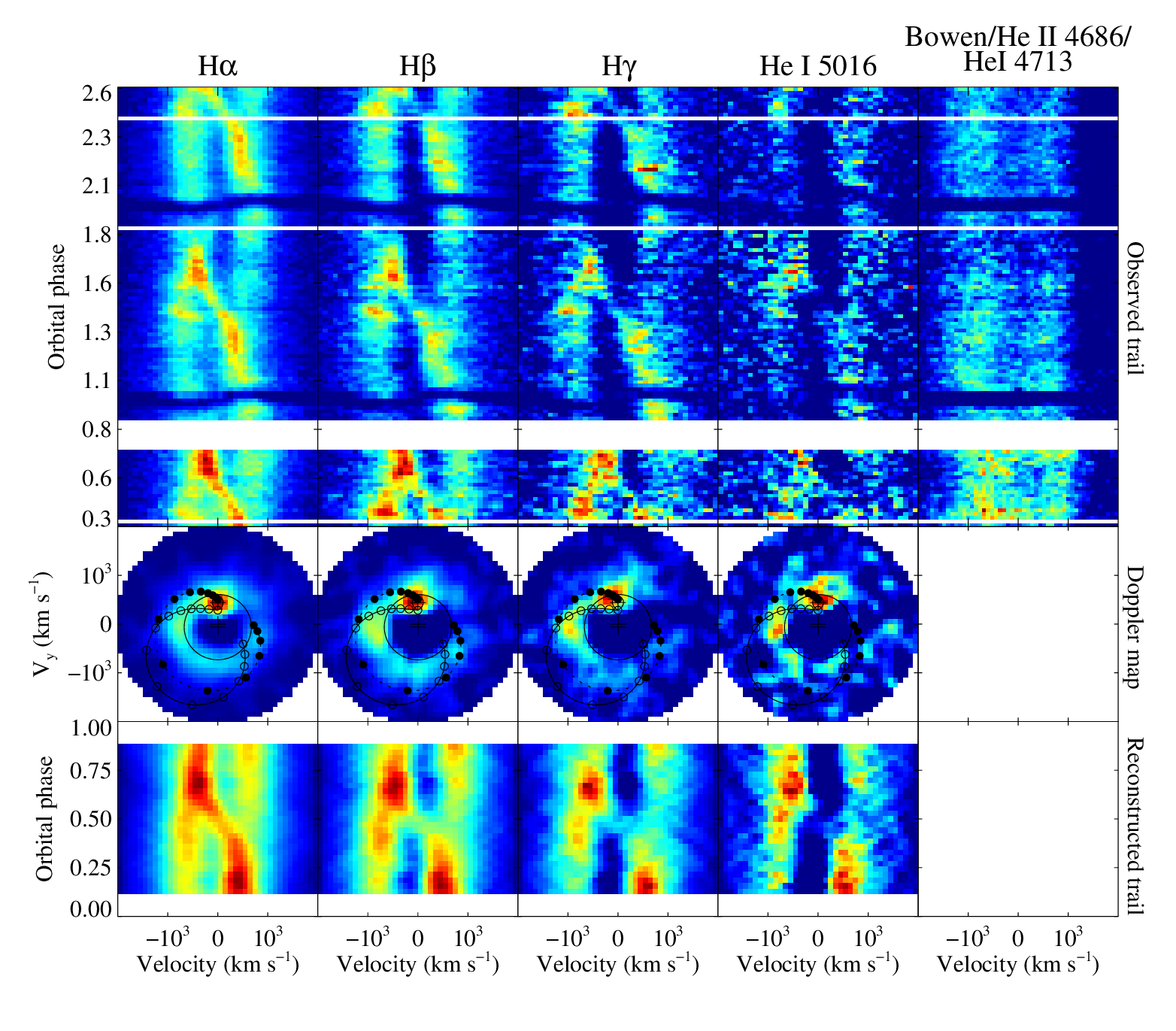,width=\textwidth}
  \caption{
    Doppler tomography during normal outburst on 7 Jan 2001. Rows as
    in Fig. \ref{Tomography34Fig}.}
  \label{Tomography7Fig}
\end{figure*}

%\begin{figure}
%  \centerline{\epsfig{file=lineeclipses7.eps,width=0.8\columnwidth}}
%  \caption{
%    Phase-folded, phase-binned lightcurves for 7 Jan 2001 \Halpha,
%    \Hbeta\ and \HeII\ for 5 separate velocity bins. Each lightcurve
%    is normalized to have average value 1 in phase range 0.1--0.2.}
%  \label{LineEclipses7Fig}
%\end{figure}

%Fig.~\ref{Tomography7Fig} shows trails and tomograms.
%{\sout{On 7 Jan, the trailed spectrum of \HeII, the Bowen blend and \HeI\ 
%4713\AA\ (last column, Fig. \ref{Tomography7Fig}) shows double-peaked
%disc emission in \HeII, but the eclipses barely move blue-red compared
%to the other lines where there is a clear blue-to-red eclipse.} } 
%Fig.  \ref{LineEclipses7Fig} shows phase-folded, phase-binned lightcurves
%for \Halpha, \Hbeta\ and \HeII\ in five separate velocity bins
%covering the line profile. {\sout{These show clearly how} The Balmer line
%eclipse is earlier in the blue peak and later in the red peak than the
%\HeII\ eclipse, while the eclipses are centred around phase 0 in the
%wings of all three lines. The ingress and egress of the \HeII\ eclipse
%also appears sharper for \HeII\ than the Balmer lines. This {\sout{is what we
%expect if}} suggests the \HeII\ emission comes from a region of the disc
%concentrated closer to the white dwarf than the Balmer emission.  
%
The
%{\sout{trailed spectra of the} 
Balmer trails (Fig. \ref{Tomography7Fig}) reveal
more complicated structure than on 2001 Jan 6
%{\sout{ thanks to the}} 
because the absorption cores are substantially
weaker. The behaviour is stable throughout the night
and is the same for all four lines, except that the core is deeper for
higher-order lines. There are double peaks and the clear blue-to-red
eclipse. There is a clear S-wave with velocity semi-amplitude
$\sim$300--600\,\kms, maximum redshift around phase 0.25, disappearing
from about phase -0.2 to 0.2.  
%{\sout{These facts suggest}} 
This S-wave
appears to originate on the inner face of the donor star. 
%{\sout{whose}}
The velocity, phasing and invisibility at
%{\sout{$\approx$440\,\kms\ and phasing would be the same as this S-wave, and
%which would be invisible in the phase range about}} 
phase -0.2 to 0.2
all suggest this, 
%{\sout{This
%feature has also been seen repeatedly in}} 
c.f. IP\,Peg during outburst
e.g. \nocite{MoralesRuedaEt:2000}Morales-Rueda, Marsh \&  Billington (2000).  

%{\sout{There is one other feature in}}
The trails suggest another S-wave. In the blue disc wing around
phase 0.4 centred on a velocity of about -700\kms, we see a bright
region which covers a phase range of less than 0.1, and a velocity
range of more than 500\kms.  This is seen in all three orbits in
\Halpha\ and \Hbeta, and can also be seen more faintly in \Hgamma\ and
\HeI. In the first orbit in the red wing at phase 0.9 we see a similar
feature at velocity of about 700\kms, probably the opposite extremum of
the same S-wave, but much fainter, c.f. OY Car
\nocite{HarlaftisMarsh:1996b}(Harlaftis \& Marsh 1996).  
%{\sout{Doppler mapping is needed to help
%further disentangle this detailed structure. These features look very
%much like those modeled in 
%, where
%outburst emission in OY\,Car was modeled with a component from the
%donor star, and with the higher velocity S-wave coming from the
%accretion stream.}}

The middle row of Fig. \ref{Tomography7Fig} shows the Doppler maps for
Jan 7. 
%{\color{red}\sout{The map for \HeII, the Bowen blend and \HeI\ 
%    4713\AA\ (last column) is distorted because it comes from the
%    blend of several lines, but the strongest component of the blend
%    is \HeII\ so this dominates the structure in the map. It reveals a
%    strong ring of emission with peak velocity $\sim1000$\,\kms, the
%    Keplerian velocity at about half the tidal truncation radius of
%    the disc, a higher velocity than the emission from the other
%    lines. This is consistent with the conclusion from the equivalent
%    widths (Section \ref{Lightcurves2001}) and the \HeII\ eclipse that
%    the \HeII\ emission comes from closer to the white dwarf than the
%    other lines. This is expected, since the hotter, more intensely
%    radiated regions are where we expect to find the excited \HeII\ 
%    ions. We must be cautious, however, since the blending of the
%    Bowen blend and \HeI\ 4713\AA\ with \HeII\ will tend to increase
%    the ring velocity in the Doppler map.}} 
The Balmer and \HeI\ 
5016\AA\ maps show strong emission concentrated around the velocity of
the donor star, 
confirming our inference from the trails and less directly from the
maps from the previous night.
%{\sout{corresponding to the S-wave observed in the trails
%which is eclipsed between phases -0.2 and 0.2, suggesting that it
%comes from the inner face of the donor star. We also detected emission
%from the donor star velocity in the non-axisymmetric components of the
%Doppler maps from Jan 6.}} 
Given the resolution, 
%{\sout{the position in the Doppler maps of}} 
this source is consistent with
emission from the inner face of the donor.  Emission from the donor
star during outburst has been seen before, e.g. OY\,Car
\nocite{HarlaftisMarsh:1996b}(Harlaftis \& Marsh 1996) and IP\,Peg
\nocite{Steeghs:2001,MoralesRuedaEt:2000,MarshHorne:1990}(Steeghs 2001; Morales-Rueda {et~al.} 2000; Marsh \& Horne 1990).  
%{\sout{ Evidence for the presence of 
%EUV emission and X-ray emission from the boundary
%layer (BL) in these systems was provided by the detection of the}}
Irradiation-driven \HeII\ 4686\AA\ emission was present in these systems,
as in IY\,UMa (Fig.~\ref{Tomography7Fig}). 
\nocite{MarshHorne:1990}Marsh \& Horne (1990) and \nocite{HarlaftisMarsh:1996b}Harlaftis \& Marsh (1996) concluded
that EUV emission and X-ray emission from the boundary
layer (BL)
could drive the Balmer emission. The 
%{\sout{strength of}}
Balmer emission in IY\,UMa is correlated with \HeII 
~(Fig.~\ref{LightcurvesFig2001}),
%{\sout{ as seen in the lightcurves}} 
suggesting the same driving mechanism.  

The other notable
feature seen in the  Jan 7 maps is an arc of emission on the left hand
side (negative $V_x$) at velocities corresponding to the outer part of
the disc. We do not see the full circle corresponding to the whole
disc.  The arcs are brightest around $V_y=0$. In \Halpha\ and \Hbeta\ 
the arc stretches through about 270$^\circ$ (measured anti-clockwise
from the donor velocity), while in \Hgamma\ and \HeI\ it only reaches
about 135$^\circ$, although \HeI\ shows a bright blob around
225$^\circ$.  This structure appears similar to that seen in outburst
in OY\,Car by \nocite{HarlaftisMarsh:1996b}Harlaftis \& Marsh (1996), which followed and was
thus attributed to the accretion stream. However, in IY\,UMa this is
not the case; it follows the velocity of the outer disc and no
stream-disc impact hotspot is seen, so this emission does not arise
from dissipation of energy where the stream hits the disc edge.
Despite the lack of emission from dissipation of kinetic energy at the
stream-disc impact, some vertically extended structure is likely to be
present where the stream impacts the disc, providing a region where we
might see reprocessing of the EUV and X-rays from the BL, simply
because vertically extended structure will intercept more BL
radiation. This would be  top left in the Doppler maps.
If this vertical structure, raised at this point initially by the
stream disc impact, extended further around the disc, it could explain
the brightest region of the emission line arcs seen here. The inner
face of this raised region should appear bright, being directly
irradiated, explaining why the corresponding feature in the
trailed spectrogram is brighter when we look at the inner face (phase
$\sim0.4$) than the outer face (phase $\sim 0.9$). Simulations of
discs and the stream-disc impact do show vertical structure extended
around the disc edge e.g.  \nocite{ArmitageLivio:1996,HiroseEt:1991}Armitage \& Livio (1996); Hirose, Osaki \& Mineshige (1991)
and there is much observational evidence for vertical structure in the
outer disc and a flared disc during outburst
\nocite{BillingtonEt:1996,IoannouEt:1999,NaylorEt:1987}(e.g. Billington {et~al.} 1996; Ioannou {et~al.} 1999; Naylor {et~al.} 1987).
If the core absorption in the lines is due to
  material in the outer disc along the line of sight to the white
  dwarf, then the weakening of the core absorption on the Jan 7
%th of January 
points towards changes in the vertical structure of the
  outer disc. Such changes might reveal vertical structure around the
  hotspot on the 
Jan 7
%th of January 
which was entirely swamped by a uniformly
  thick outer disc on Jan 6. We expect some reprocessing of
the light from the BL throughout the disc, which could explain why the
ring extends most, if not all, of the way around the disc in the
\Halpha\ and \Hbeta\ maps. We conclude that the structure in the Jan 7
maps can be explained by reprocessing of boundary layer EUV and X-ray
emission on the inner face of the donor and vertical structure in the
disc near to and perhaps triggered by the stream-disc impact.

The reconstructed Balmer and \HeI\ trails are consistent with those
observed, showing the double peaks, deep core and donor S-wave. The
feature around phase 0.4 in the blue wing is smeared out but
identifiable.

\subsection{2000 January: superoutburst}
\begin{figure*}
  \hspace*{-1.5cm}\epsfig{file=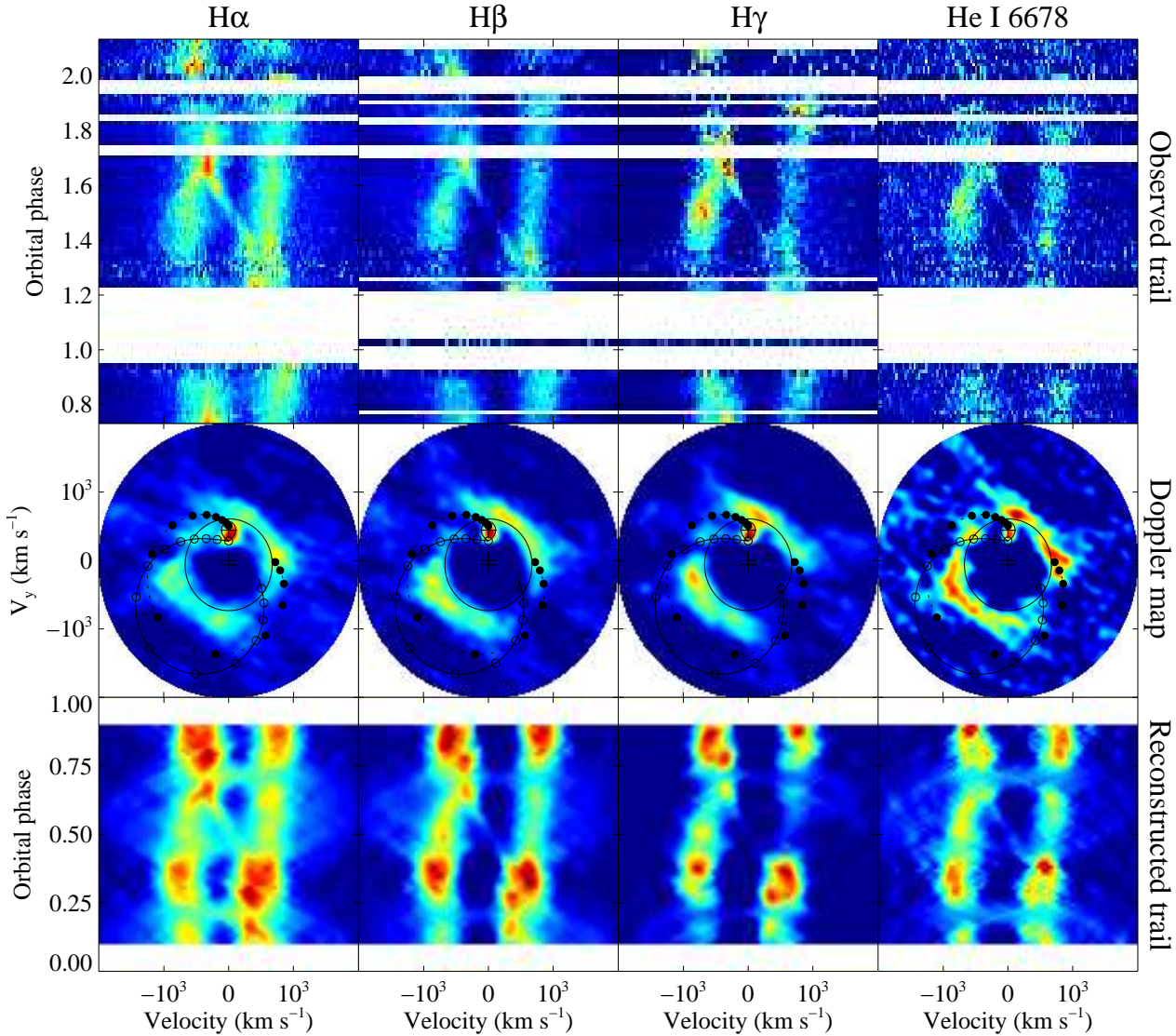,width=0.89\textwidth}
  \caption{
    Doppler tomography during superoutburst on 19 Jan 2000. Rows as
    in Fig. \ref{Tomography34Fig}.}
  \label{TomographysuperFig}
\end{figure*}

We normalised the line profiles
by scaling each
  so that the integrated line flux above the continuum
  was the same for all. This reduced the 
  variability in the flux caused by poor observing conditions.
These normalised
trailed spectra for 2000 Jan 19 
%{\sout{(during the superoutburst) are shown in the}} 
(Fig.  \ref{TomographysuperFig}, top row)
show the double-peaked structure characteristic of accretion discs.
Also present in all lines 
%{\color{red}\sout{apart from \HeII}} 
is the
deep core between the disc peaks, which is below the continuum
except in
\Halpha. The Balmer lines and \HeI\ 6678\AA\ show a
narrow partial S-wave that moves from red (phase 1.3)
to blue (phase 1.7) and
disappears around phase 1.8. This looks like the donor star emission
seen during the normal outburst but not during quiescence. 
%{\sout{The fact that}} 
We see the inner face of the donor star only during the two 
bright states, supporting the idea that 
%{\sout{ what we are seeing}}
the donor
star is being
irradiated.
%{\sout{ by the high energy emission from the boundary
%layer. The}} 
The 
%Balmer and \HeI\ 
Doppler maps 
%{\sout{ are presented in the}} 
(middle row of Fig.
\ref{TomographysuperFig})
all show disc
emission with asymmetric artefacts like those on 2001 Jan 6, which
resulted
from incomplete phase coverage. The donor emission is visible in
these 
maps.
%{\sout{lines, as in the normal outburst.}} {\color{red}\sout{ There is
%    higher-velocity disc emission in \HeII, seen most strongly in the
%    lower left quadrant of the map, with no evidence of donor
%    emission.  The higher velocity of the \HeII\ emission suggests
%    that it come from closer to the WD than the Balmer or \HeI, as
%    discussed for the normal outburst.}} 
The asymmetry probably
results 
%partly 
from the incomplete phase coverage, 
%although we also
%expect to see significant asymmetry in the disc. 
%{\sout{This is because}}
though during superoutburst the disc is tidally distorted and precessing, and
so asymmetric dissipation patterns and vertical structure in the disc
are expected \nocite{FoulkesEt:2004}(e.g. Foulkes {et~al.} 2004). 
%We therefore expect
%asymmetry 
%{\sout{ to be manifest in the emission lines,}} 
%whether lines are
%dissipation-powered or due to reprocessing.
The
reconstructed Balmer and \HeI\ trails show bright red-shifted emission
around phase 0.3 and blue-shifted emission around 0.8. This is the
donor S-wave crossing the disc peaks, and is not seen in the observed
trails because the inner face of the donor will be only half visible
at phases 0.25 and 0.75, and is completely invisible around phase 0.

\section{Discussion}

There are several sets of photometry of eclipsing dwarf novae
beginning early during the rise to outburst. \nocite{Vogt:1983b}Vogt (1983)
presents the rise to a normal outburst of OY\,Car, while
\nocite{WebbEt:1999}Webb {et~al.} (1999) presents an entire outburst of IP\,Peg.
\nocite{IoannouEt:1999}Ioannou {et~al.} (1999) presents observations of another short period
high-inclination dwarf nova, HT\,Cas, covering the rise, peak and
decline from a normal outburst.  \nocite{Vogt:1983b}Vogt (1983) and
\nocite{RuttenEt:1992}Rutten {et~al.} (1992) showed that during the rise to outburst in
OY\,Car, the outer disc rapidly brightens (as it enters the hot
ionized state), with this hot bright region quickly propagating
inwards towards the white dwarf, the outer radius remaining constant.
As the outburst reaches maximum brightness, the radius of the emission
region shrinks by about a third.  \nocite{IoannouEt:1999}Ioannou {et~al.} (1999) find exactly
the same outside-in behaviour in HT\,Cas, additionally concluding that
the disc becomes vertically flared during outburst. 
%{\sout{ We note that}}
In
IY\,UMa 
%{\sout{ we see the}}
 the emission region shrinks and the
eclipse depth increases
between the observed continuum maximum and a day later. This 
%{\sout{is a sign that}} 
indicates
the outer disc returns to quiescence before the inner
disc, as the cooling wave propagates inwards. The disc emission region
shrinks during decline in most (if not all) dwarf novae, including two
dwarf novae where eclipses suggested an outburst beginning in the
inner disc and propagating outwards \nocite{WebbEt:1999,BaptistaEt:2001}(IP\,Peg and
EX\,Dra, Webb {et~al.} 1999; Baptista \& Catal\'{a}n 2001, respectively). Eclipse
lightcurves at the very start of the outburst are required to
distinquish between inside-out and outside-in outbursts from single
band photometry alone.

The apparent disappearance of the orbital hump in \Halpha\ on 4 Jan
2001 immediately preceding the normal outburst while there is still
hotspot line emission seen in the Doppler map is curious. It must
result from a change in geometry of the hotspot making it equally
visible at all phases. This may simply be due to the inherent
variability in the stream-disc impact and disc structure, but we note
that while the amplitude of the hotspot is variable, none of our other
observations show it completely disappearing. The disappearance may be
due to a change in the structure of the outer disc as the outburst
begins. This would require the outburst to begin in the outer disc, so
that there has been no increase in luminosity detected when it occurs.

The delay in the rise of the line emission compared to the rise of the
continuum is something which can only be detected in rare (and
fortuitous) spectrophotometry in the earliest stages of outburst like
those analysed here. The delay is easily understood if the line
emission in outburst is powered by irradiation from the BL and the
outburst is of the outside-in type seen in OY\,Car and HT\,Cas. The
viscous dissipation-powered continuum begins to rise as soon as the
outer disc enters the high state, while the emission lines do not
become significantly stronger until the heating wave has reached the
inner disc 
%increasing the mass accretion rate and thus 
the intensity
of high energy BL emission and the emission lines it powers. This
delay of the lines relative to the continuum is analogous to the UV
delay in dwarf nova outburst observations \nocite{WheatEt:2003}(Wheatley, Mauche \& Mattei 2003), 
where the UV emission (from
the inner disc) rises after the visual emission which comes
predominantly from the outer disc. The strong \HeII\ emission, thought
to be powered by irradiation, the Balmer lightcurves mimicking that of
\HeII, the emission from the inner face of the donor star, and the
phase-dependent arc of emission in the 2001 Jan 7 lines all support
the conclusion that the emission lines are predominantly powered by
irradiation during outbursts. 
%{\sout{Carrying out 3D hydrodynamical
%simulations of outbursting accretion disc in IY\,UMa like those of
%\nocite{TrussEt:2000} (), and applying a prescription for irradiation by
%the white dwarf/boundary layer, it should be possible to simulate line
%profile variations and Doppler maps which can be compared with the
%observations presented here. This would provide a valuable, detailed
%test for this model and the simulations.}}

\label{discussion}

\section{Summary}

Extensive spectroscopic observations of IY\,UMa have revealed
``classic'' quiescent accretion flow, with 

\begin{itemize}
\item strong continuum and line emission from the hotspot region
\item a blackbody hotspot temperature of $\sim$11000\,K
\item ionization of hydrogen in the hotspot region suppressing Balmer emission from the disc close to the hotspot.
\end{itemize}

Rare spectrophotometry of the rise to a normal outburst reveals
\begin{itemize}
\item the continuum and line emission regions shrink at
the
  outburst peak
\item the rise in emission line flux is delayed relative to the continuum
by a few hours
  to two days 
\item during outburst Balmer, \HeII\ and Bowen blend emission is seen
  from the disc, while Balmer emission is also seen from the inner
  face of the donor and possible vertical structure in the outer disc.
\end{itemize}
We conclude that the 2001 Jan outburst was of the outside-in type,
beginning in the outer disc and propagating inwards. Irradiation by
EUV and X-rays from the boundary layer powers the emission lines
during outburst, with the time taken for mass to move inwards during
the outburst explaining the delay between the rise of the continuum
and emission line flux. 
%{\sout{ We shall test this hypothesis using
%hydrodynamical simulations in a future paper.}}

\section{Acknowledgements}

The data presented here have been taken using ALFOSC, which is owned
by the Instituto de Astrofisica de Andalucia (IAA) and operated at the
Nordic Optical Telescope under agreement between IAA and the NBIfAFG
of the Astronomical Observatory of Copenhagen. The Nordic Optical
Telescope is operated on the island of La Palma jointly by Denmark,
Finland, Iceland, Norway, and Sweden, in the Spanish Observatorio del
Roque de los Muchachos of the Instituto de Astrofisica de Canarias.
The William Herschel telescope is operated on the island of La Palma
by the Isaac Newton Group in the Spanish Observatorio del Roque de los
Muchachos of the Instituto de Astrof\'{\i}sica de Canarias. DJR was
supported by a PPARC studentship and the OU research committee and by
a PPARC rolling grant at Leicester. CAH gratefully acknowledges
support from the Leverhulme Trust F/00-180/A. LMR was supported by a
PPARC post-doctoral grant. The authors gratefully acknowledge the work
of the Variable Star Network (VSNET, {\tt
  http://www.kusastro.kyoto-u.ac.jp/vsnet/index.html}). The reduction
and analysis of the WHT data were carried out on the Southampton node
of the STARLINK network. The authors thank Rob Hynes for useful
comments.

%\bibliographystyle{natbib-mn}
%% \bibliography

\label{lastpage}

\end{document}